\documentclass[prd,twocolumn,superscriptaddress,showpacs,amssymb,amsmath,amsfonts,aps,altaffilletter]{revtex4}
\pdfoutput=1

\usepackage{color}
\usepackage{times}
\usepackage{graphicx}
\usepackage{fancyhdr}
\usepackage{float}
\usepackage{ulem}
\usepackage[raggedright]{subfigure}
\usepackage{acronym, multirow}

\makeatletter
\makeatletter
\def\@fnsymbol#1{\ifcase#1\or * \or  $+$ \or  \$ \or \#  \or \dag \or \ddag \or
$\mathsection$ \or $ \mathparagraph$ \or $\|$  \or \textordfeminine \or \textbul
let   
\or ** \or $++$ \or  \$\$ \or \#\#  \or \dag\dag \or \ddag\ddag \or
$\mathsection\mathsection$ \or $ \mathparagraph\mathparagraph$ \or $\|\|$  \or 
\textordfeminine\textordfeminine \or \textbullet \textbullet \or *** \or $+++$ 
\or  \$\$\$ \or \#\#  \or \dag\dag \or \ddag\ddag \or
$\mathsection \mathsection\mathsection$ \or $ \mathparagraph 
\mathparagraph\mathparagraph$ \or $\|\|\|$  \or 
\textordfeminine\textordfeminine\textordfeminine \or 
\textbullet\textbullet\textbullet \or \else \@ctrerr\fi}
\makeatother

\newcommand\fake[1]{\textcolor{red}{#1}}

\def\thercsid{\relax}
\def\rcsid#1{\def\next##1#1{\def\thercsid{##1}}\next}

\rcsid$Id: s5-2yr-ligovirgo-lowcbc.tex,v 1.52 2010/05/11 22:23:49 vaulin Exp $

\renewcommand{\today}{\number\day\space\ifcase\month\or
  January\or February\or March\or April\or May\or June\or
  July\or August\or September\or October\or November\or December\fi
  \space\number\year}

\def\Msun{\ensuremath{M_{\odot}}}

\def\BNSCumLum{\ensuremath{370}}
\def\BNSCalError{\ensuremath{13\%}}
\def\BNSMCError{\ensuremath{17\%}}
\def\BNSWaveformError{\ensuremath{19\%}}
\def\BNSGalDistError{\ensuremath{-16\%}}
\def\BNSGalMagError{\ensuremath{29\%}}
\def\BNSNonSpinUL{\ensuremath{8.7 \times 10^{-3} }}
\def\BNSSpinUL{...}

\def\BHNSCumLum{\ensuremath{1600}}
\def\BHNSCalError{\ensuremath{14\%}}
\def\BHNSMCError{\ensuremath{17\%}}
\def\BHNSWaveformError{\ensuremath{18\%}}
\def\BHNSGalDistError{\ensuremath{-13\%}}
\def\BHNSGalMagError{\ensuremath{30\%}}
\def\BHNSNonSpinUL{\ensuremath{2.2 \times 10^{-3} }}
\def\BHNSSpinUL{\ensuremath{2.7 \times 10^{-3} }}

\def\BBHCumLum{\ensuremath{8300}}
\def\BBHCalError{\ensuremath{14\%}}
\def\BBHMCError{\ensuremath{18\%}}
\def\BBHWaveformError{\ensuremath{16\%}}
\def\BBHGalDistError{\ensuremath{-13\%}}
\def\BBHGalMagError{\ensuremath{31\%}}
\def\BBHNonSpinUL{\ensuremath{4.4 \times 10^{-4} }}
\def\BBHSpinUL{\ensuremath{5.3 \times 10^{-4} }}

\begin{document}

\acrodef{BBH}{binary black holes}
\acrodef{BNS}{binary neutron stars}
\acrodef{BHNS}{black hole--neutron star binaries}
\acrodef{SNR}{signal-to-noise ratio}
\acrodef{SPA}{stationary-phase approximation}
\acrodef{LIGO}{Laser Interferometer Gravitational-wave Observatory}
\acrodef{LHO}{LIGO Hanford Observatory}
\acrodef{LLO}{LIGO Livingston Observatory}
\acrodef{LSC}{LIGO Scientific Collaboration}
\acrodef{GRB}{gamma-ray bursts}
\acrodef{CBC}{compact binary coalescence}
\acrodef{GW}{gravitational wave}
\acrodef{ISCO}{innermost stable circular orbit}
\acrodef{FAR}{false alarm rate}
\acrodef{IFAR}{inverse false alarm rate}
\acrodef{CL}{confidence level}
\acrodef{PN}{post-Newtonian}
\acrodef{DQ}{data quality}
\acrodef{S5YR1}{S5 first year search}
\acrodef{S5YR2A}{S5 second year, pre VSR1}
\acrodef{VSR1}{first Virgo science run}
\acrodef{IFO}{interferometer}

\title{Search for Gravitational Waves from Compact Binary Coalescence in LIGO and Virgo Data from S5 and VSR1\\
}

\affiliation{Albert-Einstein-Institut, Max-Planck-Institut f\"ur Gravitationsphysik, D-14476 Golm, Germany$^\ast$}
\affiliation{Albert-Einstein-Institut, Max-Planck-Institut f\"ur Gravitationsphysik, D-30167 Hannover, Germany$^\ast$}
\affiliation{Andrews University, Berrien Springs, MI 49104 USA$^\ast$}
\affiliation{AstroParticule et Cosmologie (APC), CNRS: UMR7164-IN2P3-Observatoire de Paris-Universit\'e Denis Diderot-Paris 7 - CEA : DSM/IRFU, France$^\dagger$}
\affiliation{Australian National University, Canberra, 0200, Australia$^\ast$}
\affiliation{California Institute of Technology, Pasadena, CA  91125, USA$^\ast$}
\affiliation{California State University Fullerton, Fullerton CA 92831 USA$^\ast$}
\affiliation{Caltech-CaRT, Pasadena, CA  91125, USA$^\ast$}
\affiliation{Cardiff University, Cardiff, CF24 3AA, United Kingdom$^\ast$}
\affiliation{Carleton College, Northfield, MN  55057, USA$^\ast$}
\affiliation{Charles Sturt University, Wagga Wagga, NSW 2678, Australia$^\ast$}
\affiliation{Columbia University, New York, NY  10027, USA$^\ast$}
\affiliation{European Gravitational Observatory (EGO), I-56021 Cascina (PI), Italy$^\dagger$}
\affiliation{Embry-Riddle Aeronautical University, Prescott, AZ   86301 USA$^\ast$}
\affiliation{E\"otv\"os University, ELTE 1053 Budapest, Hungary$^\ast$}
\affiliation{Hobart and William Smith Colleges, Geneva, NY  14456, USA$^\ast$}
\affiliation{$^a$INFN, Sezione di Firenze, I-50019 Sesto Fiorentino, Italy$^\dagger$\\$^{17b}$Universit\`a degli Studi di Urbino 'Carlo Bo', I-61029 Urbino, Italy$^\dagger$}
\affiliation{INFN, Sezione di Genova;  I-16146  Genova, Italy$^\dagger$}
\affiliation{$^a$INFN, Sezione di Napoli, I-80126 Napoli, Italy$^\dagger$\\$^{19b}$Universit\`a di Napoli 'Federico II' Complesso Universitario di Monte S.Angelo, I-80126 Napoli, Italy$^\dagger$\\$^{19c}$Universit\`a di Salerno, Fisciano, I-84084 Salerno, Italy$^\dagger$}
\affiliation{$^a$INFN, Sezione di Perugia, I-06123 Perugia, Italy$^\dagger$\\$^{20b}$Universit\`a di Perugia, I-06123 Perugia, Italy$^\dagger$}
\affiliation{$^a$INFN, Sezione di Pisa, I-56127 Pisa, Italy$^\dagger$\\$^{21b}$Universit\`a di Pisa, I-56127 Pisa, Italy$^\dagger$\\$^{21c}$Universit\`a di Siena, I-53100 Siena, Italy$^\dagger$}
\affiliation{$^a$INFN, Sezione di Roma, I-00185 Roma, Italy$^\dagger$\\$^{22b}$Universit\`a 'La Sapienza', I-00185  Roma, Italy$^\dagger$}
\affiliation{$^a$INFN, Sezione di Roma Tor Vergata, I-00133 Roma, Italy$^\dagger$\\$^{23b}$Universit\`a di Roma Tor Vergata, I-00133 Roma, Italy$^\dagger$\\$^{23c}$Universit\`a dell'Aquila, I-67100 L'Aquila, Italy$^\dagger$}
\affiliation{Institute of Applied Physics, Nizhny Novgorod, 603950, Russia$^\ast$}
\affiliation{Inter-University Centre for Astronomy and Astrophysics, Pune - 411007, India$^\ast$}
\affiliation{$^a$LAL, Universit\'e Paris-Sud, IN2P3/CNRS, F-91898 Orsay, France$^\dagger$\\$^{26b}$ESPCI, CNRS,  F-75005 Paris, France$^\dagger$}
\affiliation{Laboratoire d'Annecy-le-Vieux de Physique des Particules (LAPP), Universit\'e de Savoie, CNRS/IN2P3, F-74941 Annecy-Le-Vieux, France$^\dagger$}
\affiliation{Leibniz Universit\"at Hannover, D-30167 Hannover, Germany$^\ast$}
\affiliation{LIGO - California Institute of Technology, Pasadena, CA  91125, USA$^\ast$}
\affiliation{LIGO - Hanford Observatory, Richland, WA  99352, USA$^\ast$}
\affiliation{LIGO - Livingston Observatory, Livingston, LA  70754, USA$^\ast$}
\affiliation{LIGO - Massachusetts Institute of Technology, Cambridge, MA 02139, USA$^\ast$}
\affiliation{Laboratoire des Mat\'eriaux Avanc\'es (LMA), IN2P3/CNRS, F-69622 Villeurbanne, Lyon, France$^\dagger$}
\affiliation{Louisiana State University, Baton Rouge, LA  70803, USA$^\ast$}
\affiliation{Louisiana Tech University, Ruston, LA  71272, USA$^\ast$}
\affiliation{McNeese State University, Lake Charles, LA 70609 USA$^\ast$}
\affiliation{Montana State University, Bozeman, MT 59717, USA$^\ast$}
\affiliation{Moscow State University, Moscow, 119992, Russia$^\ast$}
\affiliation{NASA/Goddard Space Flight Center, Greenbelt, MD  20771, USA$^\ast$}
\affiliation{National Astronomical Observatory of Japan, Tokyo  181-8588, Japan$^\ast$}
\affiliation{$^a$Nikhef, National Institute for Subatomic Physics, P.O. Box 41882, 1009 DB Amsterdam, The Netherlands$^\dagger$\\$^{41b}$VU University Amsterdam, De Boelelaan 1081, 1081 HV Amsterdam, The Netherlands$^\dagger$}
\affiliation{Northwestern University, Evanston, IL  60208, USA$^\ast$}
\affiliation{$^a$Universit\'e Nice-Sophia-Antipolis, CNRS, Observatoire de la C\^ote d'Azur, F-06304 Nice, France$^\dagger$\\$^{43b}$Institut de Physique de Rennes, CNRS, Universit\'e de Rennes 1, 35042 Rennes, France$^\dagger$}
\affiliation{$^a$INFN, Gruppo Collegato di Trento, Trento, Italy$^\dagger$\\$^{44b}$Universit\`a di Trento,  I-38050 Povo, Trento, Italy$^\dagger$\\$^{44c}$INFN, Sezione di Padova, I-35131 Padova, Italy$^\dagger$\\$^{44d}$Universit\`a di Padova, I-35131 Padova, Italy$^\dagger$}
\affiliation{$^a$IM-PAN, 00-956 Warsaw, Poland$^\dagger$\\$^{45b}$Warsaw University, 00-681 Warsaw, Poland$^\dagger$\\$^{45c}$Astronomical Observatory of Warsaw University, 00-478 Warsaw, Poland$^\dagger$\\$^{45d}$CAMK-PAN, 00-716 Warsaw, Poland$^\dagger$\\$^{45e}$Bia{\l}ystok University, 15-424 Bia{\l}ystok, Poland$^\dagger$\\$^{45f}$IPJ, 05-400 \'Swierk-Otwock, Poland$^\dagger$\\$^{45g}$Institute of Astronomy, 65-265 Zielona G\'ora, Poland$^\dagger$}
\affiliation{Rochester Institute of Technology, Rochester, NY  14623, USA$^\ast$}
\affiliation{Rutherford Appleton Laboratory, HSIC, Chilton, Didcot, Oxon OX11 0QX United Kingdom$^\ast$}
\affiliation{San Jose State University, San Jose, CA 95192, USA$^\ast$}
\affiliation{Sonoma State University, Rohnert Park, CA 94928, USA$^\ast$}
\affiliation{Southeastern Louisiana University, Hammond, LA  70402, USA$^\ast$}
\affiliation{Southern University and A\&M College, Baton Rouge, LA  70813, USA$^\ast$}
\affiliation{Stanford University, Stanford, CA  94305, USA$^\ast$}
\affiliation{Syracuse University, Syracuse, NY  13244, USA$^\ast$}
\affiliation{The Pennsylvania State University, University Park, PA  16802, USA$^\ast$}
\affiliation{The University of Melbourne, Parkville VIC 3010, Australia$^\ast$}
\affiliation{The University of Mississippi, University, MS 38677, USA$^\ast$}
\affiliation{The University of Sheffield, Sheffield S10 2TN, United Kingdom$^\ast$}
\affiliation{The University of Texas at Austin, Austin, TX 78712, USA$^\ast$}
\affiliation{The University of Texas at Brownsville and Texas Southmost College, Brownsville, TX  78520, USA$^\ast$}
\affiliation{Trinity University, San Antonio, TX  78212, USA$^\ast$}
\affiliation{Tsinghua University, Beijing 100084 China$^\ast$}
\affiliation{Universitat de les Illes Balears, E-07122 Palma de Mallorca, Spain$^\ast$}
\affiliation{University of Adelaide, Adelaide, SA 5005, Australia$^\ast$}
\affiliation{University of Birmingham, Birmingham, B15 2TT, United Kingdom$^\ast$}
\affiliation{University of Florida, Gainesville, FL  32611, USA$^\ast$}
\affiliation{University of Glasgow, Glasgow, G12 8QQ, United Kingdom$^\ast$}
\affiliation{University of Maryland, College Park, MD 20742 USA$^\ast$}
\affiliation{University of Massachusetts - Amherst, Amherst, MA 01003, USA$^\ast$}
\affiliation{University of Michigan, Ann Arbor, MI  48109, USA$^\ast$}
\affiliation{University of Minnesota, Minneapolis, MN 55455, USA$^\ast$}
\affiliation{University of Oregon, Eugene, OR  97403, USA$^\ast$}
\affiliation{University of Rochester, Rochester, NY  14627, USA$^\ast$}
\affiliation{University of Salerno, I-84084 Fisciano (Salerno), Italy$^\ast$ and INFN (Sezione di Napoli), Italy$^\ast$}
\affiliation{University of Sannio at Benevento, I-82100 Benevento, Italy and INFN (Sezione di Napoli), Italy$^\ast$}
\affiliation{University of Southampton, Southampton, SO17 1BJ, United Kingdom$^\ast$}
\affiliation{University of Strathclyde, Glasgow, G1 1XQ, United Kingdom$^\ast$}
\affiliation{University of Western Australia, Crawley, WA 6009, Australia$^\ast$}
\affiliation{University of Wisconsin--Milwaukee, Milwaukee, WI  53201, USA$^\ast$}
\affiliation {Washington State University, Pullman, WA 99164, USA$^\ast$}
\author{J.~Abadie$^\text{29}$}\noaffiliation\author{B.~P.~Abbott$^\text{29}$}\noaffiliation\author{R.~Abbott$^\text{29}$}\noaffiliation\author{Abernathy$^\text{66}$}\noaffiliation\author{T.~Accadia$^\text{27}$}\noaffiliation\author{F.~Acernese$^\text{19a,19c}$}\noaffiliation\author{C.~Adams$^\text{31}$}\noaffiliation\author{R.~Adhikari$^\text{29}$}\noaffiliation\author{P.~Ajith$^\text{29}$}\noaffiliation\author{B.~Allen$^\text{2,78}$}\noaffiliation
% ...

\collaboration{$^\ast$The LIGO Scientific Collaboration and $^\dagger$The Virgo Collaboration}
\noaffiliation

\date[\relax]{ RCS \thercsid; compiled \today }
\fake{}

\begin{abstract}\quad

We report the results of the first search for gravitational waves from compact binary coalescence using data from the \ac{LIGO} and Virgo detectors. Five months of data were collected during the concurrent S5 (LIGO) and VSR1 (Virgo) science runs. The search focused on signals from binary mergers with a total mass between $2$ and $35~\Msun$. No gravitational waves are identified. The cumulative 90\%-confidence upper limits on the rate of compact binary coalescence are calculated for non-spinning binary neutron stars, black hole-neutron star systems, and binary black holes to be \BNSNonSpinUL $\textrm{ yr}^{-1} \mathrm{L}_{10}^{-1}$, \BHNSNonSpinUL $\textrm{ yr}^{-1} \mathrm{L}_{10}^{-1}$, and \BBHNonSpinUL $\textrm{ yr}^{-1} \mathrm{L}_{10}^{-1}$ respectively, where $\mathrm{L}_{10}$ is $10^{10}$ times the blue solar luminosity. These upper limits are compared with astrophysical expectations. 
\end{abstract}


\maketitle

\section{Introduction}\label{sec:overview}

The coalescence of a stellar mass compact binary is expected to produce gravitational waves detectable by ground-based interferometers. Binary neutron stars (BNS), \ac{BBH} and \ac{BHNS} can spiral together to produce signals in the frequency band where the Laser Interferometer Gravitational-wave Observatory (LIGO) \cite{Abbott:2007kv} and Virgo \cite{Acernese:2008b} detectors are most sensitive (40--1000 Hz).

LIGO was collecting data at the Hanford, Washington and Livingston, Louisiana sites as part of its fifth science run (S5) (4 November 4 2005 -- 30 September 2007) when the first science run (VSR1) began at the Virgo detector in Cascina, Italy on 18 May 2007. During VSR1 the Virgo detector operated at reduced sensitivity since its commissioning was still incomplete. LIGO data collected before 18 May 2007 were analyzed separately and upper limits on the rate of gravitational waves from binary inspirals were reported in Refs.~\cite{LIGOS3S4all,Collaboration:2009tt,Abbott:2009qj}.

Here we describe the results of the first joint search for gravitational waves from compact binary coalescence with LIGO and Virgo data. This search covers gravitational waves from binaries with a total mass between $2~\Msun$ and $35~\Msun$ and a minimum component mass of $1~\Msun$. This analysis is based on the same methods as the S5 LIGO-only searches~\cite{Collaboration:2009tt,Abbott:2009qj}. Since the analysis is considered integral in preparing for future joint searches in LIGO and Virgo data, further developments were performed to integrate Virgo into the pipeline, even though VSR1 data had limited sensitivity when compared with LIGO's S5 data. No gravitational-wave signal is identified and upper limits are calculated.

In section \ref{sec:dq}, we describe the data used in this analysis. The data reduction pipeline is explained in section \ref{sec:pipeline} and ends with a description of the detection statistic. The results and upper limits appear in sections \ref{sec:results} and \ref{sec:ul}. Details of a self-imposed blind injection challenge are given in Appendix \ref{appendix:bic}.

\section{Data Quality}\label{sec:dq}

\begin{figure}[ht]
  \includegraphics[width=3in]{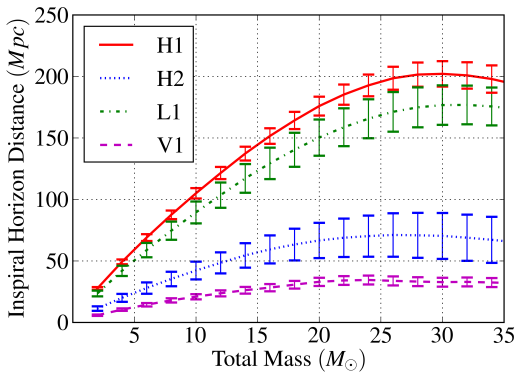}
  \caption{The average inspiral horizon distance over the run is shown as a function of the total mass of the binary system for each interferometer. The error bars indicate variation over the duration of the run.}
  \label{fig:distvsmass}
\end{figure}

The detectors are referred to as H1 (Hanford 4 km), H2 (Hanford 2 km), L1 (Livingston 4 km), and V1 (Virgo 3 km). Data from LIGO and Virgo are recorded in the same format, making it easier to run the LIGO pipeline on the additional detector. The relative sensitivities of these detectors can be assessed with horizon distance, the distance at which an optimally located, optimally oriented binary would produce triggers with a signal-to-noise-ratio (SNR) of 8 in the detector. When averaged over the duration of the search, the horizon distances for a 1.4, $1.4~\Msun$ BNS system are approximately 37, 16, 32, and 8 Mpc for H1, H2, L1, and V1 respectively. See Figure \ref{fig:distvsmass} for the horizon distance in each interferometer as a function of the total mass of the binary system. 

The detectors are very sensitive to their environments and fall out of science mode when disturbed, meaning that they are temporarily not recording science-quality data. Because the data streams from each detector are not continuous, different combinations of detectors may be taking data at any given time. As we describe in section \ref{sec:pipelineoverview}, we require time coincidence to identify possible gravitational waves and hence we only analyze the data when at least two detectors are operating. There are eleven combinations for what we define as \textit{analysis time}: H1H2, H1L1, H1V1, H2L1, H2V1, L1V1, H1H2L1, H1H2V1, H1L1V1, H2L1V1, H1H2L1V1. Analysis time indicates that the listed detectors are collecting science-quality data. Because H1 and H2 are co-located, correlated noise leads to poor background estimates and hence H1H2 time was rejected. 

A number of quality criteria were established before and during the run to reject times when the data are unreliable, either due to instrumental problems or external factors. See Appendix A of Ref.~\cite{Collaboration:2009tt} for a more thorough description of the \textit{veto categories} we use. We do not analyze data rejected by Category 1 vetoes because it indicates severe problems. Category 2 vetoes remove artifacts with well understood origin and coupling. Category 3 vetoes are based on statistical correlations, and Category 4 vetoes are least serious and only used in the candidate follow up procedure. We make our candidate event list and perform the upper limit calculation with data that passes Category 1+2+3 vetoes. We also look for loud candidates with significantly low false alarm probability that occur during times rejected by Category 3 (but that pass Category 1+2 vetoes). When only Category 1+2 vetoes are applied, 115.2 days of data are analyzed; when Category 1+2+3 vetoes are applied, 101.1 days of data are analyzed. Our two most sensitive detectors (H1 and L1) were simultaneously running during 68\% of this time. 

\section{The Data Analysis Pipeline}\label{sec:pipeline}

The data processing is performed in a similar manner to the S5 LIGO-only analyses~\cite{Collaboration:2009tt,Abbott:2009qj}, although the addition of Virgo to the pipeline led to enhancements in the ranking method for candidates. Due to long-term variations in detector performance, data are analyzed in one-month blocks of time in order to obtain more accurate background estimates. There are four approximately 30-day blocks and one 19-day block and each time period is analyzed with an identical pipeline. The results of these five periods are combined with previous analyses into one set of upper limit statements.

\subsection{Overview of Pipeline}\label{sec:pipelineoverview}

As described in the LIGO-only searches~\cite{Collaboration:2009tt}, the analysis begins with four separate data streams, one from each detector. We construct template banks~\cite{hexabank} of non-spinning post-Newtonian waveforms~\cite{Blanchet:1996pi,Droz:1999qx,Blanchet:2002av,Buonanno:2006ui,Boyle:2007ft,Hannam:2007ik,pan:024014,Boyle:2009dg,thorne.k:1987,SathyaDhurandhar:1991,Owen:1998dk}. These templates cover a range of binary mass combinations, ($m_1$, $m_2$). The single-detector data are match filtered with the templates and the resulting triggers pass to the next pipeline stage if they exceed an SNR of 5.5~\cite{Allen:2005fk}. Because the background does not follow a Gaussian distribution, the false alarm rate (FAR) is quite high in single-detector data. To reject noise artifacts, we use signal-based vetoes~\cite{LIGOS3S4Tuning,Rodriguez:2007}, including a $\chi^2$ test~\cite{Allen:2004} and require triggers from different detectors to be coincident in time and mass parameters~\cite{Robinson:2008}. We define \textit{event type} as the combination of detectors contributing to a given coincident trigger. A double coincidence trigger can occur during double, triple, or quadruple analysis time, while a quadruple coincidence can only occur during quadruple analysis time. We apply consistency tests on the coincident triggers. For example, since H1 is about twice as sensitive as H2, any coincidence that includes an H2 trigger, but not an H1 trigger when H1 was collecting data, is rejected. The remaining triggers are ranked based on an estimate of their likelihood of being a true signal or background. Any candidate events that stand out significantly above the background are followed up with a more detailed study of the triggers and detector conditions at the time of the event~\cite{detection-checklist-GWDAW07}.  

The background for the search is estimated by time-shifting the data from the different detectors. The time shifts are larger than the light-travel time between any pair of detectors, therefore any observed coincidences in this data are accidental. The L1 and V1 data streams are shifted in increments of 5 and 15 seconds, respectively, while the H1 and H2 data streams are held fixed with respect to each other. This is because H1 and H2 are co-located, and noise from environmental disturbances is correlated in these interferometers. For this same reason, the background for H1H2 triggers can not be reliably estimated. H1H2 triggers are excluded from the calculation of the upper limit, but the loudest are followed up to ensure exceptional candidates are not missed.

\subsection{Parameter Choices and Tuning}

Many analysis parameters are determined at the onset of the analysis based on known properties of the individual detectors. LIGO data is analyzed above 40 Hz, and templates for the LIGO detectors cover a region with total masses between $2~\Msun$ and $35~\Msun$. Virgo data quality information is best in the high frequency region, therefore the low frequency cutoff is set to 60 Hz for Virgo data. Consequently, the Virgo template bank is constructed to cover only the BNS mass region, with a minimum total mass of $2~\Msun$ and maximum chirp mass of $2.612~\Msun$ (where chirp mass is $\mathcal{M}_\mathrm{c}=((m_1m_2)^3/(m_1+m_2))^{1/5}$ and $m_1 = m_2 = 3~\Msun$).

When optimally tuned, veto cuts and consistency tests remove a significant number of background triggers while having minimal effect on the detection efficiency for simulated signals. With the addition of a fourth detector, the tuning was revisited. In the process of tuning we set the appropriate parameters for Virgo and verified that the corresponding parameters for the LIGO detectors did not need to be changed from those used in S5 LIGO-only analyses.

\subsection{Detection Statistic}
\label{sec:Statistic}
In Refs.~\cite{Collaboration:2009tt,Abbott:2009qj}, coincident triggers that survived all veto cuts and consistency tests \cite{LIGOS3S4Tuning} are ranked according to their combined effective SNR, $\rho_\mathrm{c}$, first used as detection statistic in the analysis of data from the S3 and S4 LIGO science runs~\cite{LIGOS3S4all}. The combined effective SNR statistic is based on the standard SNR, but it incorporates the value of the $\chi^2$ test into its definition~\cite{Allen:2004}. Its effect is to assign a lower detection statistic to those coincident triggers that have high values of the $\chi^2$, indicating that they are less consistent with the expected gravitational waveform. Further details concerning construction of the combined effective SNR can be found in Appendix C of Ref.~\cite{Collaboration:2009tt}.

As observed in previous LIGO analyses, both the total rate of triggers and their distribution over effective SNR vary strongly with total mass. Variation also exists for each event type across different analysis times. Additionally, one should consider significant differences in detector sensitivities, for example the H2 and V1 detectors are much less sensitive than either H1 or L1. As a result, some analysis times are more efficient in detecting gravitational waves than others. Within a specific analysis time, certain event types are more likely to be associated with a gravitational-wave event. Hence we specifically distinguish all of the possible combinations of event type in analysis times. 

In order to account for variation in background rates and differences in the sensitivity of the detectors, we implemented the following post-processing algorithm. First, coincident triggers that survive the main pipeline are clustered such that only the trigger with the highest combined effective SNR within a 10 s window is kept. Then clustered triggers are subdivided into categories by analysis time, event type, and mass. Based on regions of similar background behavior, we define three mass bins: $0.87 \leq \mathcal{M}_\mathrm{c}/\Msun < 3.48$, $3.48 \leq \mathcal{M}_\mathrm{c} < 7.4$, and $7.4 \leq \mathcal{M}_\mathrm{c} < 15.24$. These correspond to equal mass binaries with total masses of 2$~\Msun$ -- 8$~\Msun$, 8$~\Msun$ -- 17$~\Msun$ and 17$~\Msun$ -- 35$~\Msun$. For every trigger in each category, using our estimate of the background (time-shifted data), we calculate the rate, $R_0(\rho_\mathrm{c}, m, \alpha, \beta)$, of background triggers with combined effective SNR greater than or equal to that of the trigger. The mass bin is indicated by $m$, while $\alpha$ and $\beta$ are the event type and analysis time. Next, we introduce efficiency factors that estimate the probability of detecting a signal with a given combination of detectors in a specific analysis time. Virgo only has templates covering the BNS mass space, therefore in the calculation of the efficiency factors we use a population of simulated BNS gravitational-wave signals injected into the data. This procedure accounts for most of the effects introduced by variations in the detector sensitivities. Because the population of simulated signals is distributed uniformly in inverse distance, a reweighting is necessary. The efficiency factors are defined as: 

\begin{equation}
\epsilon(\alpha,\beta) = \frac{\sum_\mathrm{found} D_\mathrm{inj}^{3}}{\sum_\mathrm{all} D_\mathrm{inj}^{3}}. 
\end{equation}
The numerator is a sum of all injections found for that particular $\alpha$ and $\beta$. The denominator sums over all injected signals during a particular analysis time, $\beta$. $D_\mathrm{inj}$ is the injected distance to the binary.

Finally we define the detection (or ranking) statistic, $L$, for the search to be

\begin{equation}
\label{detection_statistic}
L(\rho_\mathrm{c}, m, \alpha, \beta) = \ln\left[\frac{\epsilon(\alpha,\beta)}{R_0(\rho_\mathrm{c}, m, \alpha, \beta)}\right].
\end{equation}

For a gravitational-wave detection, a candidate is expected to have an $L$ value significantly larger than the background.  We have tested this algorithm on simulated signals and find that it results in substantial increase in overall efficiency of the search.  

\section{Results}\label{sec:results}
A list of the loudest events is generated after Category 1+2+3 vetoes are applied. However, in order not to unnecessarily dismiss a possible detection, we also look for any loud candidates that might have occurred when a Category 3 veto was active (times that pass only Category 1+2 vetoes). Candidates from these times may still stand above the background, but must be closely studied to differentiate them from the elevated background noise that the Category 3 veto is intended to remove. 

\subsection{Results from Times that Pass Category 1+2+3 Vetoes}

\begin{figure}[ht]
\includegraphics[width=3in]{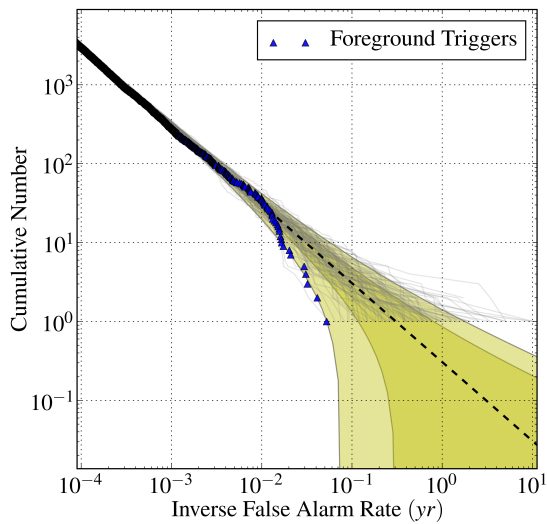}
\caption{A cumulative histogram of the inverse false alarm rate using $L$ as the detection statistic. The data are represented by the triangles and each gray line represents a background trial made from time-shifting the data against itself. The darker and lighter shaded regions denote $N^{1/2}$ and $2N^{1/2}$ errors, respectively. The data combine triggers from all five LIGO-Virgo months when Category 1+2+3 vetoes are applied.}
\label{fig:ifar}
\end{figure}

After Category 1+2+3 vetoes are applied, we find no events with a detection statistic significantly larger than the background estimation. In Figure \ref{fig:ifar}, the data are overlaid on the background. The inverse false alarm rate is calculated with detection statistic $L$ defined by Eq. (\ref{detection_statistic}). The loudest trigger in the five-month span is an H1L1 coincidence from H1H2L1V1 time with a false alarm rate of 19 per year. As 0.28 yr was searched, this is consistent with the background expectation.

However, as seen in Figure \ref{fig:ifar}, there are fewer foreground triggers than the mean background. While the foreground lies within the $2N^{1/2}$ uncertainties, we performed a series of tests to exclude an error in the analysis or a bias in the way the foreground was handled with respect to the background. We ran the analysis on simulated Gaussian noise data. No deficit of foreground triggers in the tail of the distribution was found, which suggests that there is no problem in the analysis procedure or codes. We studied how the data quality and vetoes were applied in the analysis and found no error. We also changed the segmentation of the data and observed that the foreground events were shifted within the expectation for random fluctuations. Thus, we conclude that the results are consistent with a fluctuation of the foreground compared to the background.

\subsection{Results from Times that Do Not Pass Category 3 Vetoes}

An event list was generated for times when Category 3 vetoes were active, meaning that the events were only able to survive the Category 1+2 vetoes. Only one event is inconsistent with the estimated background. That sole significant candidate, an H1H2L1 triple coincidence, is a hardware injection, part of a \textit{blind injection challenge}. During four months of S5/VSR1, the LIGO and Virgo Collaborations agreed that simulated signals would be inserted into the LIGO-only data without the search groups knowing the time or number of injections and their parameters. This was an exercise to test the effectiveness of the search procedures and all blind injection triggers are removed from the results presented in this publication.

This sole candidate corresponds to a blind injection signal that was injected into the LIGO data during a time of high seismic activity at low frequencies at the LIGO Livingston Observatory. A Category 3 veto rejected this time period and hence this blind injection signal was not identified in the Category 1+2+3 event list. Unfortunately, the parameters of the blind injection challenge were revealed after the Category 1+2+3 event list was produced, but before we looked for significant candidates that might have been removed by Category 3 vetoes. Hence, the follow up procedure for significant candidates was not exercised until after the injection parameter were known. Detailed investigations related to the blind injection challenge are described in Appendix \ref{appendix:bic}.

\subsection{Results for H1H2 Double Coincidences}

Although we do not have reliable estimates of the detection statistic for H1H2 events, we did look for interesting H1H2 candidates and found one that passed Category 1+2+3 vetoes. It corresponds to the same blind injection mentioned earlier. When the candidate was vetoed in L1, it became an H1H2 double candidate (see Appendix \ref{appendix:bic}). No other interesting H1H2 candidates are identified. 

\section{Upper Limits}\label{sec:ul}

Other than the blind injection candidate, no significant candidates are identified after Category 1+2+3 vetoes are applied or when Category 3 vetoes are disregarded. We calculate upper limits on the rate of compact binary coalescence for the following astrophysical objects after Category 1+2+3 vetoes are applied: BNS $[m_1 = m_2 = (1.35\pm 0.04)~\Msun]$, BHNS $[m_1 = (5\pm 1)~\Msun,~m_2 = (1.35\pm 0.04)~\Msun]$, and BBH $[m_1 = m_2 = (5\pm 1)~\Msun]$. We also present upper limits as a function of the total mass of the binary and as a function of the black hole mass for BHNS binaries. 

The upper limits are reported for both non-spinning and spinning objects in Table \ref{tab:ulresults}. Only non-spinning templates are used in this search, so there is an additional loss of efficiency associated with spinning waveforms that leads to slightly less-constrained upper limits in the spinning case. The results are reported as a rate in units of number per $\mathrm{L}_{10}$ per year, where $\mathrm{L}_{10}$ is $10^{10}$ times the blue solar luminosity, which is expected to be proportional to the binary coalescence rate~\cite{LIGOS3S4Galaxies}. The horizon distance listed in Table \ref{tab:ulresults} is approximated for the H1 or L1 detector and is a good estimate of the sensitivity of the search.

We calculate our upper limits using the loudest event from the search, as described in Ref.~\cite{Biswas:2007ni}. In this method, the posterior distribution for the rate depends on two quantities, ${\cal C}_{L}$ and $\Lambda$, that are functions of the loudness parameter, $x$. In our experiment, $x$ is the inverse false alarm rate of the loudest observed event according to the detection statistic in Equation \ref{detection_statistic}. $\Lambda$ is a measure of the likelihood of detecting a single event with loudness parameter, $x$, versus such an event occurring in the experimental background. ${\cal C}_{L}$ is the cumulative luminosity of sources that produce signals that are louder than $x$. Assuming a uniform prior, the posterior distribution for the rate of coalescence is given by:

\begin{equation}
\label{eqn:posterior}
p \left({ \mu | {\cal C}_{L}, T, \Lambda }\right) = \frac{{\cal C}_{L} T}{1 + \Lambda} \left({1 + \mu {\cal C}_{L} T \Lambda}\right) e^{-\mu {\cal C}_{L} T},
\end{equation}
where $\mu$ is the rate and $T$ is the analyzed time. In general, $\Lambda$ is given by~\cite{Biswas:2007ni}:
\begin{equation}
\Lambda \left({ x }\right) = \left({ - \frac{1}{{\cal C}_{L}} \frac{d {\cal C}_{L}}{dx} }\right) \left({ \frac{1}{P_0} \frac{d P_0}{dx} }\right)^{-1},
\label{eqn:lambda}
\end{equation}
where $P_0$ is the probability of obtaining zero background events louder than $x$ for the given search and observation time.

\begin{figure}
\includegraphics[width=3in]{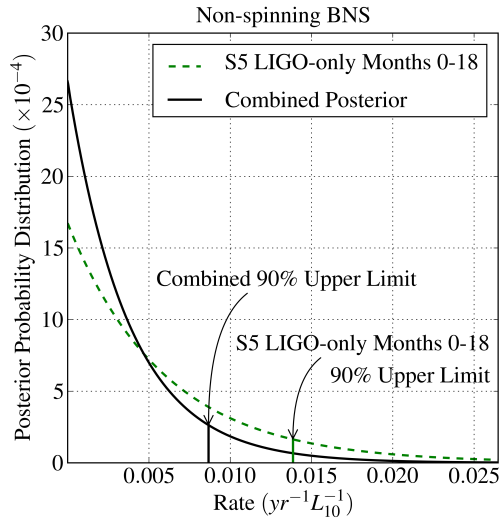}
\caption{The posterior probability distribution for the rate of non-spinning BNS coalescence. The results of all previous LIGO searches are included in the plot as the prior, labeled as S5 LIGO-only Months 0--18. Each of the five LIGO-Virgo month results was combined with the prior to obtain the combined posterior, shown as the solid black line.}
   \label{fig:ul}
\end{figure}

The cumulative luminosity measures how many potential sources we can detect with this search, based on the blue light luminosity of galaxies. To find the cumulative luminosity, we take the product of the detection efficiency, calculated as a function of mass and distance, and the luminosity from galaxies in the catalog~\cite{LIGOS3S4Galaxies} and integrate over distance. We marginalize over our uncertainties when calculating the cumulative luminosity using the values given in Table \ref{tab:ulresults}. These include detector calibration, Monte Carlo error, distances and luminosities given in the galaxy catalog and inaccuracies in the template waveforms~\cite{Fairhurst:2007qj}. The results from all five months and the prior S5 results~\cite{Collaboration:2009tt,Abbott:2009qj} are combined by taking the product of their posterior distributions calculated with uniform priors as in Equation \ref{eqn:posterior}. Figure \ref{fig:ul} shows the probability distribution from the combined data for the rate of BNS coalescence.

When spin is neglected and the priors from previous LIGO searches are used, the upper limits on the rate of compact binary coalescence are 

\begin{eqnarray}
\mathcal{R}_{90\%,{\rm BNS}} = \BNSNonSpinUL\, \mathrm{yr}^{-1}\mathrm{L_{10}}^{-1} \\
\mathcal{R}_{90\%,{\rm BHNS}} = \BHNSNonSpinUL\, \mathrm{yr}^{-1}\mathrm{L_{10}}^{-1} \\
\mathcal{R}_{90\%,{\rm BBH}} =  \BBHNonSpinUL\, \mathrm{yr}^{-1}\mathrm{L_{10}}^{-1},
\end{eqnarray}
which are consistent with upper limit estimates based solely on the sensitivity and observation time of the detectors~\cite{LIGOS3S4Galaxies}.

Astrophysical observations of neutron stars indicate that their spins will be too small to have a significant effect on \ac{BNS} waveforms observable by \ac{LIGO}~\cite{ATNF:psrcat,Apostolatos:1994}, hence we do not report upper limits for spinning BNS systems. However, we do consider spin effects on the upper limit for BHNS and BBH systems. The black hole spin, $S$, must be less than $G m^{2}/c$. We sample from a uniform distribution of possible spin values in order to simulate the effect of spin on our ability to detect the binary system. With black hole spin included, the upper limits on the rate of compact binary coalescence are

\begin{eqnarray}
\mathcal{R}_{90\%,{\rm BHNS}} = \BHNSSpinUL\, \mathrm{yr}^{-1}\mathrm{L_{10}}^{-1} \\
\mathcal{R}_{90\%,{\rm BBH}} =  \BBHSpinUL\, \mathrm{yr}^{-1}\mathrm{L_{10}}^{-1}.
\end{eqnarray}

\begin{figure}[ht]
  \includegraphics[width=3in]{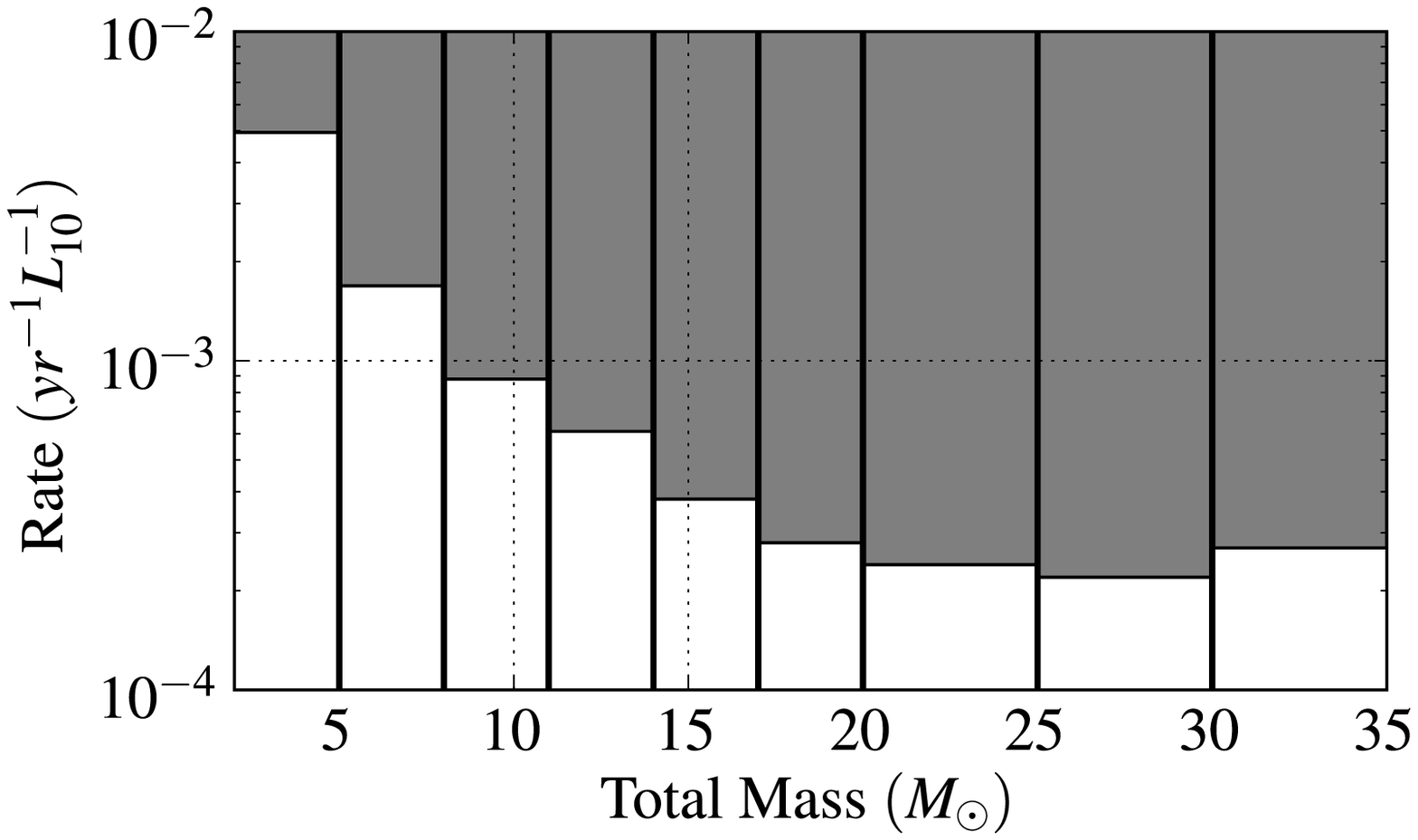}\vspace*{0.15cm}
  \includegraphics[width=3in]{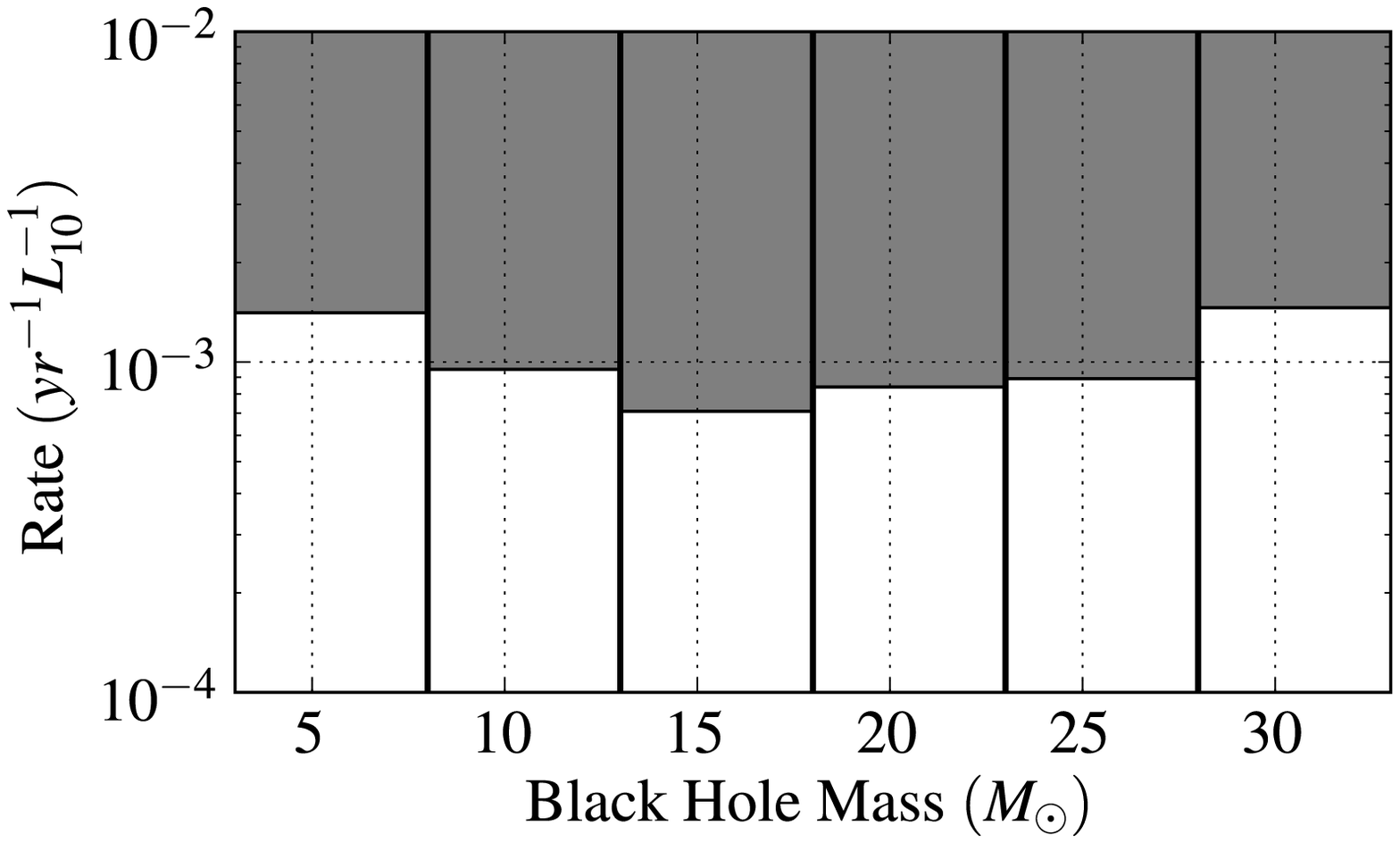}
  \caption{The 90\% rate upper limits as a function of mass. The first figure gives the upper limit on the rate of coalescence from BBH system as a function of the total mass of the system. The second figure gives the BHNS upper limit as a function of black hole mass, assuming a fixed neutron star mass of $1.35~\Msun$.}
  \label{fig:ulmass}
\end{figure}

We also produce two sets of upper limits as a function of mass. The BBH upper limit shown in Figure \ref{fig:ulmass} assumes a uniform distribution of the component mass. The BHNS upper limit is shown as a function of black hole mass, assuming a fixed neutron star mass of $1.35~\Msun$.

\begin{table*}[t]
\center
\begin{tabular}{l | c | c | c}
\hline
\hline
 & BNS & BHNS & BBH\\
\hline
Component Masses $\left(M_{\odot}\right)$ & 1.35/1.35 & 5.0/1.35 & 5.0/5.0 \\
\hline
Horizon Distance (Mpc) & $\sim$ 30 & $\sim$ 50 & $\sim$ 90 \\
\hline
Cumulative Luminosity $\left({L_{10}}\right)$ & $\BNSCumLum$ & $\BHNSCumLum$ & $\BBHCumLum$ \\
\hline
Calibration Error & $\BNSCalError$ & $\BHNSCalError$ & $\BBHCalError$ \\
\hline
Monte Carlo Error & $\BNSMCError$ & $\BHNSMCError$ & $\BBHMCError$ \\
\hline
Waveform Error & $\BNSWaveformError$ & $\BHNSWaveformError$ & $\BBHWaveformError$ \\
\hline
Galaxy Distance Error & $\BNSGalDistError$ & $\BHNSGalDistError$ & $\BBHGalDistError$ \\
\hline
Galaxy Magnitude Error & $\BNSGalMagError$ & $\BHNSGalMagError$ & $\BBHGalMagError$ \\
\hline
Non-spinning Upper Limit $\left({{\rm yr}^{-1} L_{10}^{-1}}\right)$ & $\BNSNonSpinUL$ & $\BHNSNonSpinUL$ & $\BBHNonSpinUL$ \\
\hline
Spinning Upper Limit $\left({{\rm yr}^{-1} L_{10}^{-1}}\right)$ & $\BNSSpinUL$ & $\BHNSSpinUL$ & $\BBHSpinUL$ \\
\hline
\hline
\end{tabular}
\caption{Summary of results. The horizon distance is averaged over the time of the search. The cumulative luminosity combines the detection efficiency with the galaxy catalog luminosity. Here, the value is the time-weighted average of the cumulative luminosity for each month. Many uncertainties are included in the calculation of the upper limit and they are summarized over all months. The effects of spin on BNS systems are negligible and not reported here.}
\label{tab:ulresults}
\end{table*}

\section{Conclusions}\label{sec:conclusions}

We searched for gravitational waves from compact binary coalescence in the mass region $2~\Msun$ to $35~\Msun$. Over 101 days of coincident data were collected during the end of the LIGO S5 and Virgo VSR1 runs, making this the first joint search for gravitational waves from compact binaries with LIGO and Virgo data. The LIGO data analysis pipeline was augmented to handle the extra complexity of four detectors and a larger number of coincidence categories. Although no gravitational-wave candidates are identified, upper limits on rates of binary coalescence are established. The upper limits improve when combined with the previous LIGO-only results. These upper limits are still more than an order of magnitude larger than optimistic astrophysical expectations \cite{ratesdoc}. Hardware upgrades after S5 and VSR1 completed should yield better sensitivity in future searches. With the advent of three-site analyses, sky localization techniques are being developed to reconstruct the direction of any gravitational-wave sources detected in the future.

\acknowledgments

The authors gratefully acknowledge the support of the United States
National Science Foundation for the construction and operation of the
LIGO Laboratory, the Science and Technology Facilities Council of the
United Kingdom, the Max-Planck-Society, and the State of
Niedersachsen/Germany for support of the construction and operation of
the GEO600 detector, and the Italian Istituto Nazionale di Fisica
Nucleare and the French Centre National de la Recherche Scientifique
for the construction and operation of the Virgo detector. The authors
also gratefully acknowledge the support of the research by these
agencies and by the Australian Research Council, the Council of
Scientific and Industrial Research of India, the Istituto Nazionale di
Fisica Nucleare of Italy, the Spanish Ministerio de Educaci\'on y
Ciencia, the Conselleria d'Economia Hisenda i Innovaci\'o of the
Govern de les Illes Balears, the Foundation for Fundamental Research
on Matter supported by the Netherlands Organisation for Scientific Research, 
the Polish Ministry of Science and Higher Education, the FOCUS
Programme of Foundation for Polish Science,
the Royal Society, the Scottish Funding Council, the
Scottish Universities Physics Alliance, The National Aeronautics and
Space Administration, the Carnegie Trust, the Leverhulme Trust, the
David and Lucile Packard Foundation, the Research Corporation, and
the Alfred P. Sloan Foundation. This document has been assigned LIGO Laboratory document number P0900305-v6.

\appendix

\section{Blind Injection Challenge}\label{appendix:bic}

During the \textit{blind injection challenge}, simulated signals were inserted into the LIGO-only data without the search groups knowing the time or number of injections and their parameters. Two \textit{blind injections} occurred in the data described in this paper. The first simulated a burst of gravitational waves. The injected signal was the sum of two Gaussian modulated sinusoids with linearly time-varying frequency. The root-square-sum amplitude $h_\mathrm{rss}$ for the signal was $1.0 \times 10^{-21}$ at the Earth. The dominant component was at 58 Hz and the duration was about $12$ ms. This injection was not a target of this analysis and was not identified as a significant candidate. However, see Ref.~\cite{S5VSR1Burst} about the significance of this injection in the Collaboration's burst search. 

The second blind injection was the simulated binary inspiral signal referred to in Section \ref{sec:results}. The waveform simulated a binary system with masses $1.1$ and $5.1~\Msun$, with small spins 0.19 and 0.06, respectively, in dimensionless units of the spin parameter $\hat{a} = \left({c S}\right) / \left({G m^2}\right)$, at effective distance of 34.6 Mpc for Hanford and 42.2 Mpc for Livingston. The candidate identified in H1, H2, and L1 has non-spinning templates with masses (1.0, 5.9), (1.0, 5.7), (1.1, 5.6) $\Msun$, and effective distances (43.6, 33.2, and 42.2) Mpc respectively. Given the parameters of the signal, the absence of a Virgo trigger in the coincidence does not cast any doubt on the validity of the candidate. The loudest coincidence in the time slide background had a false alarm rate of 1 per 14 years. As this candidate was louder than that, 1 per 14 years is only a bound on its significance level.

Since a candidate was identified coincident with this injection, we conducted an extensive follow up study \cite{detection-checklist-GWDAW07}. As part of the study, the SNR time series, $\chi^2$ time series and time-frequency spectrograms at the time of the candidate were inspected. We also studied environmental influences on the detectors since this candidate was vetoed by a Category 3 data quality flag produced for high seismic noise at low frequencies at the Livingston Observatory.

The 30--40 minute period of high seismic activity was due to earthquakes near Sumatra, which produced large ground motion at Livingston (but not Hanford) at frequencies between 0.03 and 0.1 Hz. There was a higher rate of accidental coincidences in the time-slid data when using L1 triggers produced a few minutes after the time of the candidate. These accidental coincidences were coincident with peaks in seismic activity and excess power in the L1 gravitational-wave channel. These accidental coincidences were H1L1 (double) coincidences, less significant than the triple coincidence candidate seen at the time of the blind injection. No triple accidental coincidences were observed within that active seismic time. The time series of the gravitational-wave channel at the time of the candidate does not bear any resemblance to those at the times of the double coincidences correlated to seismic noise.

The candidate passed all tests related to the pipeline and the statistical analysis. The presence of the seismic data quality flag in the time around this candidate does not substantially downgrade its significance. Had there not been a blind injection at the time of this candidate, it would have been recognized as having an interesting level of statistical significance, and we would likely have pursued this candidate by dismissing seismic activity as its cause. We are developing improved methods to better estimate the significance of such detection candidates.


\bibliography{../bibtex/iulpapers}

\begin{thebibliography}{30}
\expandafter\ifx\csname natexlab\endcsname\relax\def\natexlab#1{#1}\fi
\expandafter\ifx\csname bibnamefont\endcsname\relax
  \def\bibnamefont#1{#1}\fi
\expandafter\ifx\csname bibfnamefont\endcsname\relax
  \def\bibfnamefont#1{#1}\fi
\expandafter\ifx\csname citenamefont\endcsname\relax
  \def\citenamefont#1{#1}\fi
\expandafter\ifx\csname url\endcsname\relax
  \def\url#1{\texttt{#1}}\fi
\expandafter\ifx\csname urlprefix\endcsname\relax\def\urlprefix{URL }\fi
\providecommand{\bibinfo}[2]{#2}
\providecommand{\eprint}[2][]{\url{#2}}

\bibitem[{\citenamefont{Abbott et~al.}(2009{\natexlab{a}})}]{Abbott:2007kv}
\bibinfo{author}{\bibfnamefont{B.}~\bibnamefont{Abbott}} \bibnamefont{et~al.}
  (\bibinfo{collaboration}{LIGO Scientific Collaboration}),
  \bibinfo{journal}{Rept.~Prog.~Phys.} \textbf{\bibinfo{volume}{72}},
  \bibinfo{pages}{076901} (\bibinfo{year}{2009}{\natexlab{a}}),
  \eprint{arXiv:0711.3041}.

\bibitem[{\citenamefont{{Acernese} et~al.}(2008)}]{Acernese:2008b}
\bibinfo{author}{\bibfnamefont{F.}~\bibnamefont{{Acernese}}}
  \bibnamefont{et~al.}, \bibinfo{journal}{Class. Quantum Grav.}
  \textbf{\bibinfo{volume}{25}}, \bibinfo{pages}{184001}
  (\bibinfo{year}{2008}).

\bibitem[{\citenamefont{Abbott et~al.}(2008)}]{LIGOS3S4all}
\bibinfo{author}{\bibfnamefont{B.}~\bibnamefont{Abbott}} \bibnamefont{et~al.}
  (\bibinfo{collaboration}{{LIGO} Scientific Collaboration}),
  \bibinfo{journal}{Phys.~Rev.~D} \textbf{\bibinfo{volume}{77}},
  \bibinfo{pages}{062002} (\bibinfo{year}{2008}), \eprint{arXiv:0704.3368}.

\bibitem[{\citenamefont{Abbott
  et~al.}(2009{\natexlab{b}})}]{Collaboration:2009tt}
\bibinfo{author}{\bibfnamefont{B.}~\bibnamefont{Abbott}} \bibnamefont{et~al.}
  (\bibinfo{collaboration}{LIGO Scientific Collaboration}),
  \bibinfo{journal}{Phys.~Rev.~D} \textbf{\bibinfo{volume}{79}},
  \bibinfo{pages}{122001} (\bibinfo{year}{2009}{\natexlab{b}}),
  \eprint{arXiv:0901.0302}.

\bibitem[{\citenamefont{Abbott et~al.}(2009{\natexlab{c}})}]{Abbott:2009qj}
\bibinfo{author}{\bibfnamefont{B.}~\bibnamefont{Abbott}} \bibnamefont{et~al.}
  (\bibinfo{collaboration}{LIGO Scientific Collaboration}),
  \bibinfo{journal}{Phys.~Rev.~D} \textbf{\bibinfo{volume}{80}},
  \bibinfo{pages}{047101} (\bibinfo{year}{2009}{\natexlab{c}}),
  \eprint{arXiv:0905.3710}.

\bibitem[{\citenamefont{Cokelaer}(2007)}]{hexabank}
\bibinfo{author}{\bibfnamefont{T.}~\bibnamefont{Cokelaer}},
  \bibinfo{journal}{Phys.~Rev.~D} \textbf{\bibinfo{volume}{76}},
  \bibinfo{pages}{102004} (\bibinfo{year}{2007}), \eprint{arXiv:0706.4437}.

\bibitem[{\citenamefont{Blanchet et~al.}(1996)\citenamefont{Blanchet, Iyer,
  Will, and Wiseman}}]{Blanchet:1996pi}
\bibinfo{author}{\bibfnamefont{L.}~\bibnamefont{Blanchet}},
  \bibinfo{author}{\bibfnamefont{B.~R.} \bibnamefont{Iyer}},
  \bibinfo{author}{\bibfnamefont{C.~M.} \bibnamefont{Will}}, \bibnamefont{and}
  \bibinfo{author}{\bibfnamefont{A.~G.} \bibnamefont{Wiseman}},
  \bibinfo{journal}{Class. Quant. Grav.} \textbf{\bibinfo{volume}{13}},
  \bibinfo{pages}{575} (\bibinfo{year}{1996}).

\bibitem[{\citenamefont{Droz et~al.}(1999)\citenamefont{Droz, Knapp, Poisson,
  and Owen}}]{Droz:1999qx}
\bibinfo{author}{\bibfnamefont{S.}~\bibnamefont{Droz}},
  \bibinfo{author}{\bibfnamefont{D.~J.} \bibnamefont{Knapp}},
  \bibinfo{author}{\bibfnamefont{E.}~\bibnamefont{Poisson}}, \bibnamefont{and}
  \bibinfo{author}{\bibfnamefont{B.~J.} \bibnamefont{Owen}},
  \bibinfo{journal}{Phys.~Rev.~D} \textbf{\bibinfo{volume}{59}},
  \bibinfo{pages}{124016} (\bibinfo{year}{1999}).

\bibitem[{\citenamefont{Blanchet}(2002)}]{Blanchet:2002av}
\bibinfo{author}{\bibfnamefont{L.}~\bibnamefont{Blanchet}},
  \bibinfo{journal}{Living Rev. Rel.} \textbf{\bibinfo{volume}{5}},
  \bibinfo{pages}{3} (\bibinfo{year}{2002}), \eprint{arXiv:gr-qc/0202016}.

\bibitem[{\citenamefont{Buonanno et~al.}(2007)\citenamefont{Buonanno, Cook, and
  Pretorius}}]{Buonanno:2006ui}
\bibinfo{author}{\bibfnamefont{A.}~\bibnamefont{Buonanno}},
  \bibinfo{author}{\bibfnamefont{G.~B.} \bibnamefont{Cook}}, \bibnamefont{and}
  \bibinfo{author}{\bibfnamefont{F.}~\bibnamefont{Pretorius}},
  \bibinfo{journal}{Phys.~Rev.~D} \textbf{\bibinfo{volume}{75}},
  \bibinfo{pages}{124018} (\bibinfo{year}{2007}).

\bibitem[{\citenamefont{Boyle et~al.}(2007)}]{Boyle:2007ft}
\bibinfo{author}{\bibfnamefont{M.}~\bibnamefont{Boyle}} \bibnamefont{et~al.},
  \bibinfo{journal}{Phys.~Rev.~D} \textbf{\bibinfo{volume}{76}},
  \bibinfo{pages}{124038} (\bibinfo{year}{2007}), \eprint{arXiv:0710.0158}.

\bibitem[{\citenamefont{Hannam et~al.}(2008)}]{Hannam:2007ik}
\bibinfo{author}{\bibfnamefont{M.}~\bibnamefont{Hannam}} \bibnamefont{et~al.},
  \bibinfo{journal}{Phys.~Rev.~D} \textbf{\bibinfo{volume}{77}},
  \bibinfo{pages}{044020} (\bibinfo{year}{2008}), \eprint{arXiv:0706.1305}.

\bibitem[{\citenamefont{Pan et~al.}(2008)}]{pan:024014}
\bibinfo{author}{\bibfnamefont{Y.}~\bibnamefont{Pan}} \bibnamefont{et~al.},
  \bibinfo{journal}{Phys.~Rev.~D} \textbf{\bibinfo{volume}{77}},
  \bibinfo{pages}{024014} (\bibinfo{year}{2008}).

\bibitem[{\citenamefont{Boyle et~al.}(2009)\citenamefont{Boyle, Brown, and
  Pekowsky}}]{Boyle:2009dg}
\bibinfo{author}{\bibfnamefont{M.}~\bibnamefont{Boyle}},
  \bibinfo{author}{\bibfnamefont{D.~A.} \bibnamefont{Brown}}, \bibnamefont{and}
  \bibinfo{author}{\bibfnamefont{L.}~\bibnamefont{Pekowsky}},
  \bibinfo{journal}{Class.~Quantum Grav.} \textbf{\bibinfo{volume}{26}},
  \bibinfo{pages}{114006} (\bibinfo{year}{2009}), \eprint{arXiv:0901.1628}.

\bibitem[{\citenamefont{Thorne}(1987)}]{thorne.k:1987}
\bibinfo{author}{\bibfnamefont{K.~S.} \bibnamefont{Thorne}}, in
  \emph{\bibinfo{booktitle}{Three hundred years of gravitation}}, edited by
  \bibinfo{editor}{\bibfnamefont{S.~W.} \bibnamefont{Hawking}}
  \bibnamefont{and} \bibinfo{editor}{\bibfnamefont{W.}~\bibnamefont{Israel}}
  (\bibinfo{publisher}{Cambridge University Press},
  \bibinfo{address}{Cambridge}, \bibinfo{year}{1987}),
  chap.~\bibinfo{chapter}{9}, pp. \bibinfo{pages}{330--458}.

\bibitem[{\citenamefont{Sathyaprakash and
  Dhurandhar}(1991)}]{SathyaDhurandhar:1991}
\bibinfo{author}{\bibfnamefont{B.~S.} \bibnamefont{Sathyaprakash}}
  \bibnamefont{and} \bibinfo{author}{\bibfnamefont{S.~V.}
  \bibnamefont{Dhurandhar}}, \bibinfo{journal}{Phys. Rev D}
  \textbf{\bibinfo{volume}{44}}, \bibinfo{pages}{3819} (\bibinfo{year}{1991}).

\bibitem[{\citenamefont{Owen and Sathyaprakash}(1999)}]{Owen:1998dk}
\bibinfo{author}{\bibfnamefont{B.~J.} \bibnamefont{Owen}} \bibnamefont{and}
  \bibinfo{author}{\bibfnamefont{B.~S.} \bibnamefont{Sathyaprakash}},
  \bibinfo{journal}{Phys.~Rev.~D} \textbf{\bibinfo{volume}{60}},
  \bibinfo{pages}{022002} (\bibinfo{year}{1999}).

\bibitem[{\citenamefont{Allen et~al.}(2005)\citenamefont{Allen, Anderson,
  Brady, Brown, and Creighton}}]{Allen:2005fk}
\bibinfo{author}{\bibfnamefont{B.}~\bibnamefont{Allen}},
  \bibinfo{author}{\bibfnamefont{W.~G.} \bibnamefont{Anderson}},
  \bibinfo{author}{\bibfnamefont{P.~R.} \bibnamefont{Brady}},
  \bibinfo{author}{\bibfnamefont{D.~A.} \bibnamefont{Brown}}, \bibnamefont{and}
  \bibinfo{author}{\bibfnamefont{J.~D.~E.} \bibnamefont{Creighton}}
  (\bibinfo{year}{2005}), \eprint{arXiv:gr-qc/0509116}.

\bibitem[{\citenamefont{Abbott et~al.}(2007)}]{LIGOS3S4Tuning}
\bibinfo{author}{\bibfnamefont{B.}~\bibnamefont{Abbott}} \bibnamefont{et~al.}
  (\bibinfo{collaboration}{{LIGO Scientific Collaboration}}),
  \bibinfo{type}{Tech. Rep.} \bibinfo{number}{{LIGO}-T070109-01}
  (\bibinfo{year}{2007}),
  \urlprefix\url{http://www.ligo.caltech.edu/docs/T/T070109-01.pdf}.

\bibitem[{\citenamefont{Rodr\'iguez}(2007)}]{Rodriguez:2007}
\bibinfo{author}{\bibfnamefont{A.}~\bibnamefont{Rodr\'iguez}}, Master's thesis,
  \bibinfo{school}{Louisiana State University} (\bibinfo{year}{2007}),
  \eprint{arXiv:0802.1376}.

\bibitem[{\citenamefont{Allen}(2005)}]{Allen:2004}
\bibinfo{author}{\bibfnamefont{B.}~\bibnamefont{Allen}},
  \bibinfo{journal}{Phys.~Rev.~D} \textbf{\bibinfo{volume}{71}},
  \bibinfo{pages}{062001} (\bibinfo{year}{2005}).

\bibitem[{\citenamefont{Robinson et~al.}(2008)\citenamefont{Robinson,
  Sathyaprakash, and Sengupta}}]{Robinson:2008}
\bibinfo{author}{\bibfnamefont{C.~A.~K.} \bibnamefont{Robinson}},
  \bibinfo{author}{\bibfnamefont{B.~S.} \bibnamefont{Sathyaprakash}},
  \bibnamefont{and} \bibinfo{author}{\bibfnamefont{A.~S.}
  \bibnamefont{Sengupta}}, \bibinfo{journal}{Phys.~Rev.~D}
  \textbf{\bibinfo{volume}{78}}, \bibinfo{eid}{062002} (\bibinfo{year}{2008}).

\bibitem[{\citenamefont{Gouaty and the LIGO
  Scientific~Collaboration}(2008)}]{detection-checklist-GWDAW07}
\bibinfo{author}{\bibfnamefont{R.}~\bibnamefont{Gouaty}} \bibnamefont{and}
  \bibinfo{author}{\bibnamefont{the LIGO Scientific~Collaboration}},
  \bibinfo{journal}{Class. Quantum Grav.} \textbf{\bibinfo{volume}{25}},
  \bibinfo{pages}{184006} (\bibinfo{year}{2008}), \eprint{arXiv:0805.2412}.

\bibitem[{\citenamefont{Kopparapu et~al.}(2008)\citenamefont{Kopparapu, Hanna,
  Kalogera, O'Shaughnessy, Gonzalez, Brady, and Fairhurst}}]{LIGOS3S4Galaxies}
\bibinfo{author}{\bibfnamefont{R.~K.} \bibnamefont{Kopparapu}},
  \bibinfo{author}{\bibfnamefont{C.}~\bibnamefont{Hanna}},
  \bibinfo{author}{\bibfnamefont{V.}~\bibnamefont{Kalogera}},
  \bibinfo{author}{\bibfnamefont{R.}~\bibnamefont{O'Shaughnessy}},
  \bibinfo{author}{\bibfnamefont{G.}~\bibnamefont{Gonzalez}},
  \bibinfo{author}{\bibfnamefont{P.~R.} \bibnamefont{Brady}}, \bibnamefont{and}
  \bibinfo{author}{\bibfnamefont{S.}~\bibnamefont{Fairhurst}},
  \bibinfo{journal}{\apj} \textbf{\bibinfo{volume}{675}}, \bibinfo{pages}{1459}
  (\bibinfo{year}{2008}).

\bibitem[{\citenamefont{Biswas et~al.}(2009)\citenamefont{Biswas, Brady,
  Creighton, and Fairhurst}}]{Biswas:2007ni}
\bibinfo{author}{\bibfnamefont{R.}~\bibnamefont{Biswas}},
  \bibinfo{author}{\bibfnamefont{P.~R.} \bibnamefont{Brady}},
  \bibinfo{author}{\bibfnamefont{J.~D.~E.} \bibnamefont{Creighton}},
  \bibnamefont{and}
  \bibinfo{author}{\bibfnamefont{S.}~\bibnamefont{Fairhurst}},
  \bibinfo{journal}{Class. Quantum Grav.} \textbf{\bibinfo{volume}{26}},
  \bibinfo{pages}{175009} (\bibinfo{year}{2009}), \eprint{arXiv:0710.0465}.

\bibitem[{\citenamefont{Brady and Fairhurst}(2008)}]{Fairhurst:2007qj}
\bibinfo{author}{\bibfnamefont{P.~R.} \bibnamefont{Brady}} \bibnamefont{and}
  \bibinfo{author}{\bibfnamefont{S.}~\bibnamefont{Fairhurst}},
  \bibinfo{journal}{Class. Quantum Grav.} \textbf{\bibinfo{volume}{25}},
  \bibinfo{pages}{105002} (\bibinfo{year}{2008}), \eprint{arXiv:0707.2410}.

\bibitem[{\citenamefont{Manchester et~al.}(2005)\citenamefont{Manchester,
  Hobbs, Teoh, and Hobbs}}]{ATNF:psrcat}
\bibinfo{author}{\bibfnamefont{R.~N.} \bibnamefont{Manchester}},
  \bibinfo{author}{\bibfnamefont{G.~B.} \bibnamefont{Hobbs}},
  \bibinfo{author}{\bibfnamefont{A.}~\bibnamefont{Teoh}}, \bibnamefont{and}
  \bibinfo{author}{\bibfnamefont{M.}~\bibnamefont{Hobbs}},
  \bibinfo{journal}{Astronom. J.} \textbf{\bibinfo{volume}{129}},
  \bibinfo{pages}{1993} (\bibinfo{year}{2005}).

\bibitem[{\citenamefont{Apostolatos et~al.}(1994)\citenamefont{Apostolatos,
  Cutler, Sussman, and Thorne}}]{Apostolatos:1994}
\bibinfo{author}{\bibfnamefont{T.~A.} \bibnamefont{Apostolatos}},
  \bibinfo{author}{\bibfnamefont{C.}~\bibnamefont{Cutler}},
  \bibinfo{author}{\bibfnamefont{G.~J.} \bibnamefont{Sussman}},
  \bibnamefont{and} \bibinfo{author}{\bibfnamefont{K.~S.}
  \bibnamefont{Thorne}}, \bibinfo{journal}{Phys.~Rev.~D}
  \textbf{\bibinfo{volume}{49}}, \bibinfo{pages}{6274} (\bibinfo{year}{1994}).

\bibitem[{\citenamefont{Abadie et~al.}(2010{\natexlab{a}})}]{ratesdoc}
\bibinfo{author}{\bibfnamefont{J.}~\bibnamefont{Abadie}} \bibnamefont{et~al.}
  (\bibinfo{collaboration}{LIGO Scientific Collaboration and Virgo
  Collaboration}) (\bibinfo{year}{2010}{\natexlab{a}}),
  \eprint{arXiv:1003.2480}.

\bibitem[{\citenamefont{Abadie et~al.}(2010{\natexlab{b}})}]{S5VSR1Burst}
\bibinfo{author}{\bibfnamefont{J.}~\bibnamefont{Abadie}} \bibnamefont{et~al.}
  (\bibinfo{collaboration}{LIGO Scientific Collaboration and Virgo
  Collaboration}), \bibinfo{journal}{Phys. Rev.}
  \textbf{\bibinfo{volume}{D81}}, \bibinfo{pages}{102001}
  (\bibinfo{year}{2010}{\natexlab{b}}), \eprint{arXiv:1002.1036}.

\end{thebibliography}

\end{document}